\documentstyle[prl,aps,tighten,floats]{revtex}
\newbox\rotbox
\newcommand{\be}{\begin{eqnarray}}
\newcommand{\ee}{\end{eqnarray}}
\newcommand\tag{\hbox to hsize}

\def\mytoday#1{{}\ifcase\month\or
January\or February\or March\or April\or May\or June\or
July\or August\or September\or October\or November\or December\fi
 \space \number\year}

\begin{document}
%\draft
\twocolumn

\vskip 2 cm
\noindent{\it }\hfill HD-TVP 97-12

\title{Bag Models with Fuzzy Boundaries\cite{ackn} }
\author{Hilmar Forkel} 
\address{Institut f{\"u}r Theoretische Physik, Universit{\"a}t 
Heidelberg}
\address{Philosophenweg 19, D-69120 Heidelberg, Germany}
%\date{September 1997}
\date{\today}
\maketitle
\vskip -.7cm
\begin{abstract}
We discuss how hadronic bag models can be generalized in the framework  
of fuzzy set theory to implement effects of a smooth and extended 
phase boundary. 
\end{abstract}
\vskip 0.3 cm
\narrowtext
Idealizations in physical models typically arise either from 
insufficient knowledge of the underlying physics, or from the 
desire to make them more transparent and amenable to 
quantitative analysis. Hadronic bag models \cite{cho74,hos96} 
furnish a typical example for both cases. They impose 
quark confinement inside hadrons, in a region of modified vacuum, 
by static boundary conditions at a bag radius $R$. Whereas the real 
vacuum is expected to return to its normal phase outside of the 
hadron gradually, this simple prescription leads to an 
infinitely thin and energetically unfavorable bag boundary, and 
to an abrupt transition between the two phases. 

In the following we will discuss a new implementation of extended 
boundaries \cite{for299} which is easy to apply to even the most 
complex bag models. This approach is formulated in terms of fuzzy 
set theory \cite{zad65,dub80}, in which ordinary sets are generalized 
by assigning partial memberships to their elements. The application 
to the transition between the inside and outside regions of bag models 
offers itself naturally since fuzzy sets were specifically designed 
to implement smooth transitions between unrealistically distinct 
domains in mathematical models. 

The fuzzy boundary can be most easily envisioned by considering 
the sharp surface of the standard bag model at a given radius as 
the sole element of an ordinary set. By letting this set become 
fuzzy, an extended boundary -- containing conventional bag surfaces 
of varying radii and weights as elements -- emerges. In analogy with 
the boundary conditions of standard bag models, the underlying fuzzy 
set $\rho$, the ``fuzzy bag radius'', is prescribed according to general physical requirements.

Fuzzy sets \cite{zad65} consist of an ordinary reference set $\cal{X}$ 
and a real-valued membership function 
\be
\mu  : \qquad {\cal{X}} \rightarrow  [0,1] \qquad x \mapsto  \mu(x) \, ,
\ee
which specifies the degree to which an element $x \in \cal{X}$ belongs 
to $\mu$. (Following common practice, we use the same symbol for both 
the fuzzy set and its membership function.)

Accordingly, the fuzzy bag radius is represented by a membership function 
$\rho(R)$, which specifies the degree to which a sphere of radius $R$ 
belongs to the extended bag boundary. Its reference set 
$\cal{R} \subseteq [{\rm 0} , \infty]$ minimally contains the radii 
in the surface region. We denote the center (in radial direction) of 
the boundary by $R_0$ and its width by $\Delta$. Some of the potential 
of this description of the boundary derives from the fact \cite{dub80} 
that membership degrees in fuzzy sets are generally not additive (in 
contrast to, e.g., probability measures). This implies that bag surfaces 
at different $R$ can be correlated and coexisting in a common, 
extended boundary.

Since bag models do not provide any dynamics for the boundary, we 
have to rely on more general physical requirements to find the 
appropriate shape of $\rho$ \cite{for299}. The typical surface shapes
found in nontopological soliton models \cite{fri77}, in particular, do 
not show significant asymmetries between the inner and outer parts of 
the surface. This suggests the use of a Gaussian 
membership function 
\be
\rho^{(g)}(R) = \exp \left[ \frac{-(R-R_0)^2}{2 \Delta^2} 
\right] \label{gauss}
\ee
for the fuzzy bag radius, to which we will adhere below. In order to 
check the dependence of the results on the detailed shape of the 
membership function, we have also tested alternative choices such 
as the triangular form 
$\rho^{(t)}(R) = 1-\left| \frac{R_0 - R}{ 2 \Delta} \right| \; {\rm 
for} \; \left| R - R_0 \right| \le 2 \Delta$, and $\rho^{(t)}(R) = 0$ 
otherwise.  (Note that $\rho^{(t)} \subseteq \rho^{(g)}$.) In all 
cases, the standard bag model is recovered for $\Delta \rightarrow 0$. 

The next step in the setup of the fuzzy bag model deals with the 
definition and calculation of observables. Starting from a 
conventional bag model with crisp bag radius, this is accomplished by 
employing the extension principle \cite{zad75} of fuzzy set theory. 
Adapted to the present context, it states that any map $A(R)$ 
from a (crisp) bag radius $R$ to an observable $A \in {\cal{A}}$ 
(as calculated in conventional bag models) can be uniquely extended 
to a map from the fuzzy bag radius $\rho(R)$ to a fuzzy set
\be
\nu: \qquad {\cal{F}}_I ({\cal{R}})  \rightarrow  {\cal{F}} (\cal{A}), 
\qquad 
\rho(R) \mapsto  \nu_\rho(A) \nonumber \\  \nu_\rho(x) := \sup 
\left\{ \rho(R) \, | \, R \in {\cal{R}} \, \wedge \, x = A(R) \right\}. 
\label{extprinc}
\ee
Equation (\ref{extprinc}) quantifies how the fuzziness of the basic  
variable $R$ propagates into the observables. It follows directly from 
the rules which govern fuzzy sets \cite{zad75}.

In order to convert fuzzy-bag results, i.e. the fuzzy sets $\nu(A)$, 
into numerical predictions, they have to be mapped onto those real 
numbers $\tilde{A}$ which best represent their physical information 
content. For this purpose we employ the standard centroid map 
\cite{kru94} 
\be
\tilde{A} = \frac{\int A \, \nu(A) \, dA}{\int \nu(A) \, dA} .
\label{centroid}
\ee
(The integrals extend over $\cal{A}$.) In subsequent calculations, 
the fuzzy results $\nu(A)$ should be manipulated directly whenever 
the involved mathematical operations can be extended to fuzzy 
intervals. 

The above steps complete the definition of the fuzzy bag model as the 
most direct and transparent fuzzy-set extension of the standard bag 
model. We now apply these concepts to the nonlinear chiral bag model 
of Ref. \cite{hos96} and select two results which illustrate 
characteristic properties of the fuzzy extension. The bag-radius 
dependence of the total bag energy $E$ in the hedgehog state is 
indicated in Fig. 2 (dotted line). In the corresponding fuzzy bag 
model, $E(R)$ extends to the fuzzy set 
\be
\epsilon_\rho (x) = \sup \left\{ \rho(R) \, | \, R \in 
{\cal{R}}_\epsilon 
\wedge x = E(R) \right\}, \label{enfuz}
\ee
which is plotted in Fig. 1a for $R_0 = 0.7 \, {\rm fm}$, $\Delta =  0.3 
\, {\rm fm}$, and ${\cal{R}}_\epsilon = [0,1.5]\, {\rm fm}$.

As a second example, we consider the nucleon's axial coupling $g_A$ 
to first order in the angular velocity $\Omega$ \cite{hos96}. It is 
plotted in Fig. 3 (dotted line) and shows a significantly stronger 
bag-radius dependence than the hedgehog energy, which implies 
a strong deviation  from ``Cheshire-Cat'' behavior (see below). The 
corresponding fuzzy set 
\be
\gamma_\rho (x) = \sup \left\{ \rho(R) \, | \, R \in {\cal{R}}_\gamma
\wedge x = g_A (R) \right\} \label{gafuz}
\ee
is shown in Fig. 1b for ${\cal{R}}_\gamma = [0,1] \, {\rm fm}$ with 
$R_0$ and $\Delta$ as above. The shapes of $\epsilon$ and $\gamma$ 
closely reflect the behavior of $E(R)$ and $g_A(R)$ and share some 
general properties of fuzzy bag-model observables, as discussed
in Ref. \cite{for299}.

Next, we calculate the centroids of $\epsilon(E)$ and $\gamma(g_A)$ 
according to Eq. (\ref{centroid}) and examine the dependence of the 
resulting fuzzy-bag observables $\tilde{E}$ and $\tilde{g}_A$ on 
location and extension of the boundary region. Figure 2 shows the 
hedgehog energy $\tilde{E}$ as a function of $R_0$ for different 
values of the ``fuzziness'' parameter $\Delta$. (In the following, 
we drop the tilde on fuzzy-bag results and identify them by their 
$R_0$-dependence.) The dotted line corresponds to $\Delta \rightarrow 
0$.

For increasing boundary diffuseness, the bag energy becomes 
less sensitive to $R_0$ until, beyond $\Delta \sim 0.4 \, {\rm fm}$, 
it remains almost $R_0$-independent. With $\Delta = 0.4 \, 
{\rm fm}$ and for $R_0$ in the range $0 \le R_0 \le 1 {\rm fm}$, e.g., 
$g_A$ deviates less than 10 \% from its experimental value 1.26 
\cite{pdg}. Furthermore, the extended boundary shifts the minimum of 
the fuzzy-bag energy towards smaller radii, from 0.85 to 0.5 fm. 
The reduced sensitivity of fuzzy bag model results to the boundary 
position reduces the parameter dependence of the model and complies 
with the Cheshire-Cat principle \cite{rho94}. Indeed, exact Cheshire-Cat 
models are fixed points under fuzzification \cite{for299}.

In order to get an idea of the model dependence associated with 
different (e.g., triangular and Gaussian) boundary shapes, one may 
use the fuzzy measure \cite{kru94}
\be
\parallel \mu_1 = \mu_2 \parallel \; \, = \; \inf \left\{1- |\mu_1 (x) 
- \mu_2 (x)| \; \; | x \in {\cal{X}} \right\}
\ee
for the equality of two fuzzy sets $\mu_1, \mu_2$, which yields 
$\parallel \rho^{(g)} = \rho^{(t)} \parallel \; \simeq 0.9$ and 
$\parallel \epsilon^{(g)} = \epsilon^{(t)} \parallel \; \simeq 0.85$,  $\parallel \gamma^{(g)} = \gamma^{(t)} \parallel \;\simeq 0.9$ (almost independently of $\Delta$), and indicates the predictions to be 
rather robust. 

In summary, fuzzy bag models incorporate effects of a smooth phase 
boundary in a simple and rather unbiased way, while maintaining the 
appealing simplicity and the absolute confinement of conventional 
bag models. The first results are encouraging and provide motivation 
for further conceptual developments and the exploration of potential
links, e.g., to fuzzy spaces and noncommutative geometry. 
Moreover, it seems likely that interesting physical applications 
for fuzzy sets beyond the realms of the bag model and hadronic 
physics can be found. 

The author would like to thank Bart Kosko, Mannque Rho and Georges 
Ripka for helpful comments on the fuzzy bag and the organizers 
and participants of the Bled workshop for the lively atmosphere and 
discussions.

\begin{figure}[bht]
\caption{The membership function of a) the bag energy and b) the
nucleon's axial coupling ($R_0 = 0.7$ fm, $\Delta = 0.3$ fm).}
\label{fig1}
\end{figure}
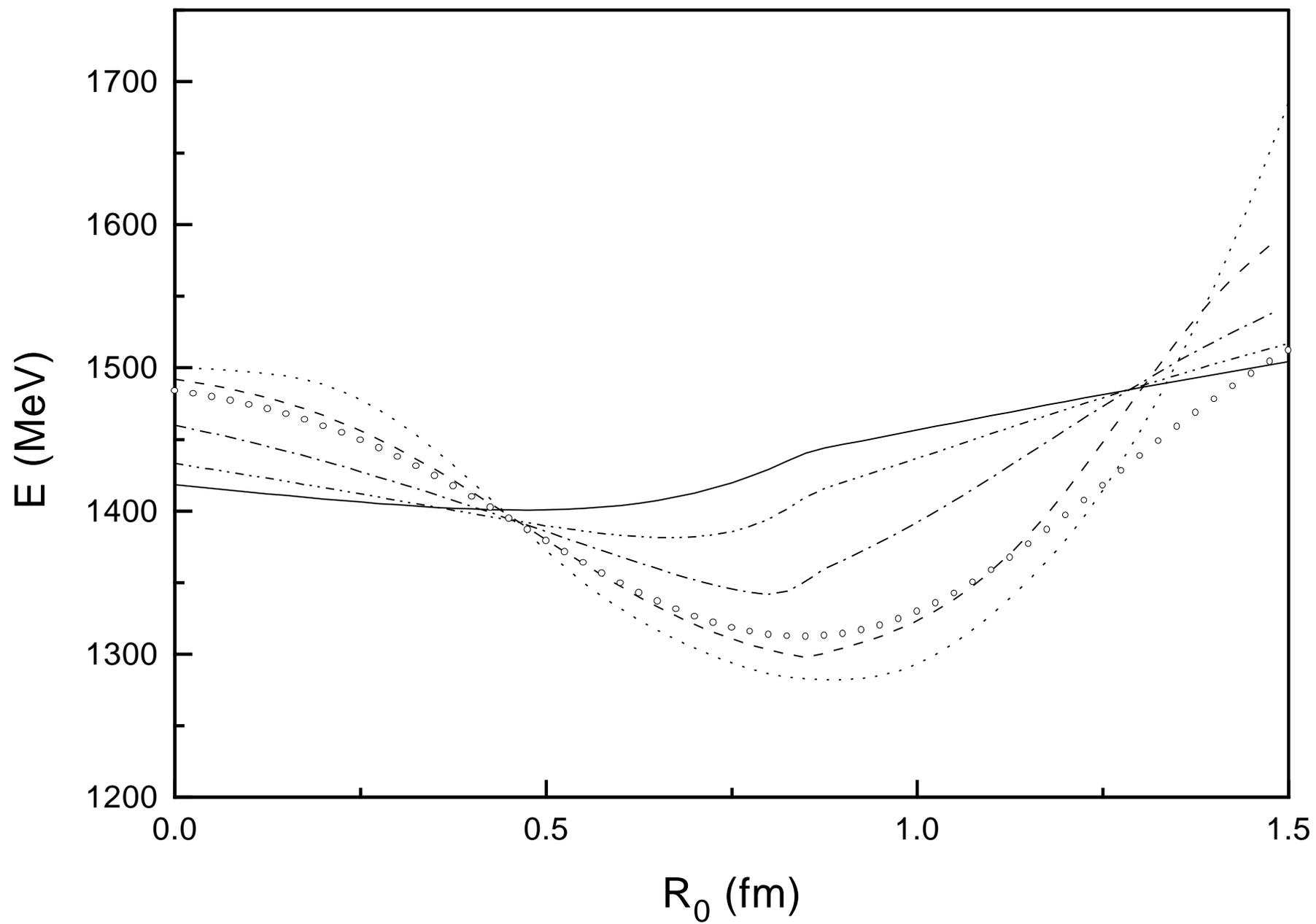
\begin{figure}[hbt]
\caption{The bag energy for $\Delta =$ 0 fm (dotted line), 0.1 fm 
(dashed), 0.2 fm (dot-dashed), 0.3 fm (dot-dot-dashed), 0.4 fm 
(solid). }
\label{fig2}
\end{figure}
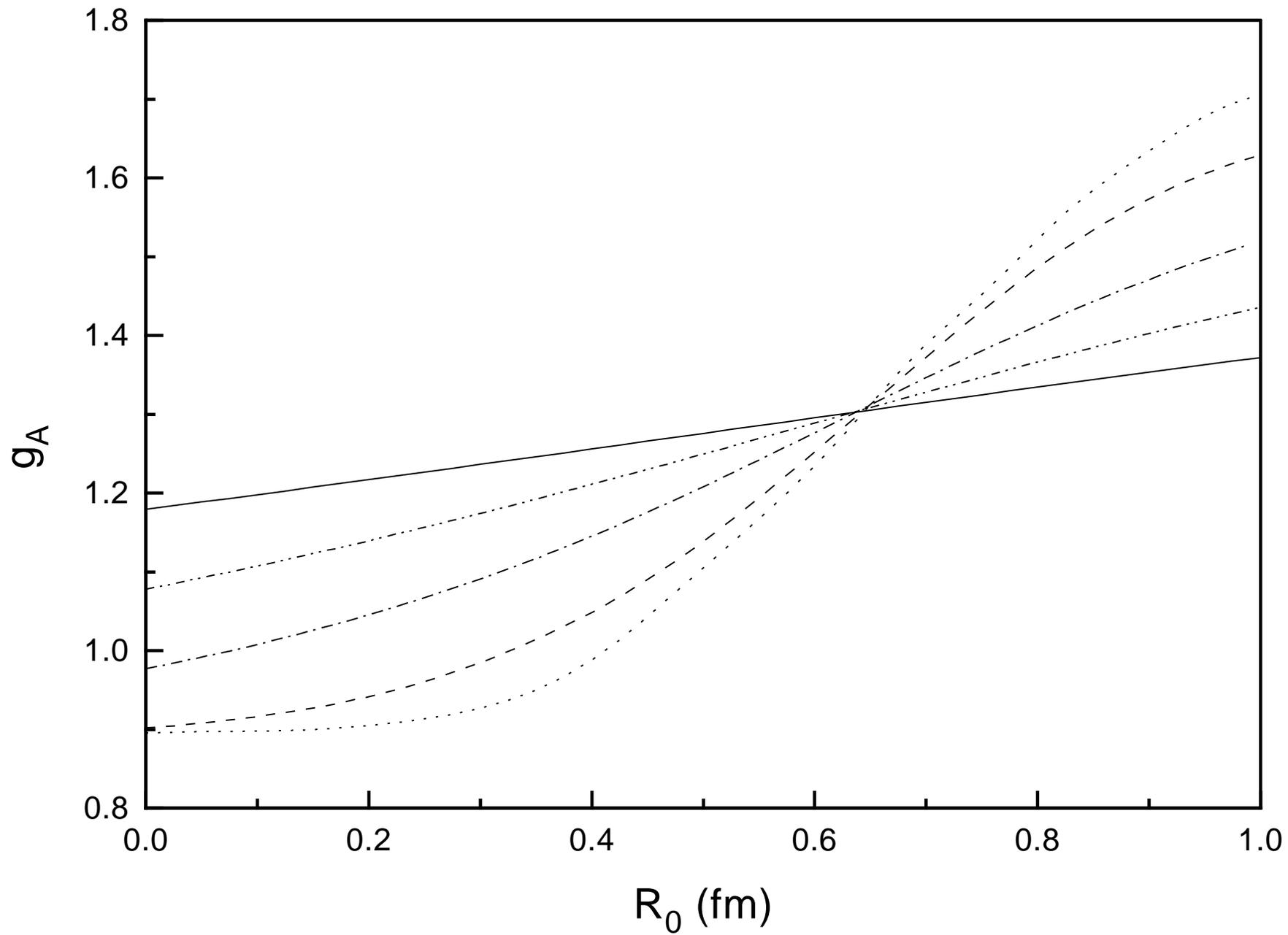
\begin{figure}[hbt]
\caption{The nucleon's axial coupling for the values of $\Delta$ 
as above. }
\label{fig3}
\end{figure}

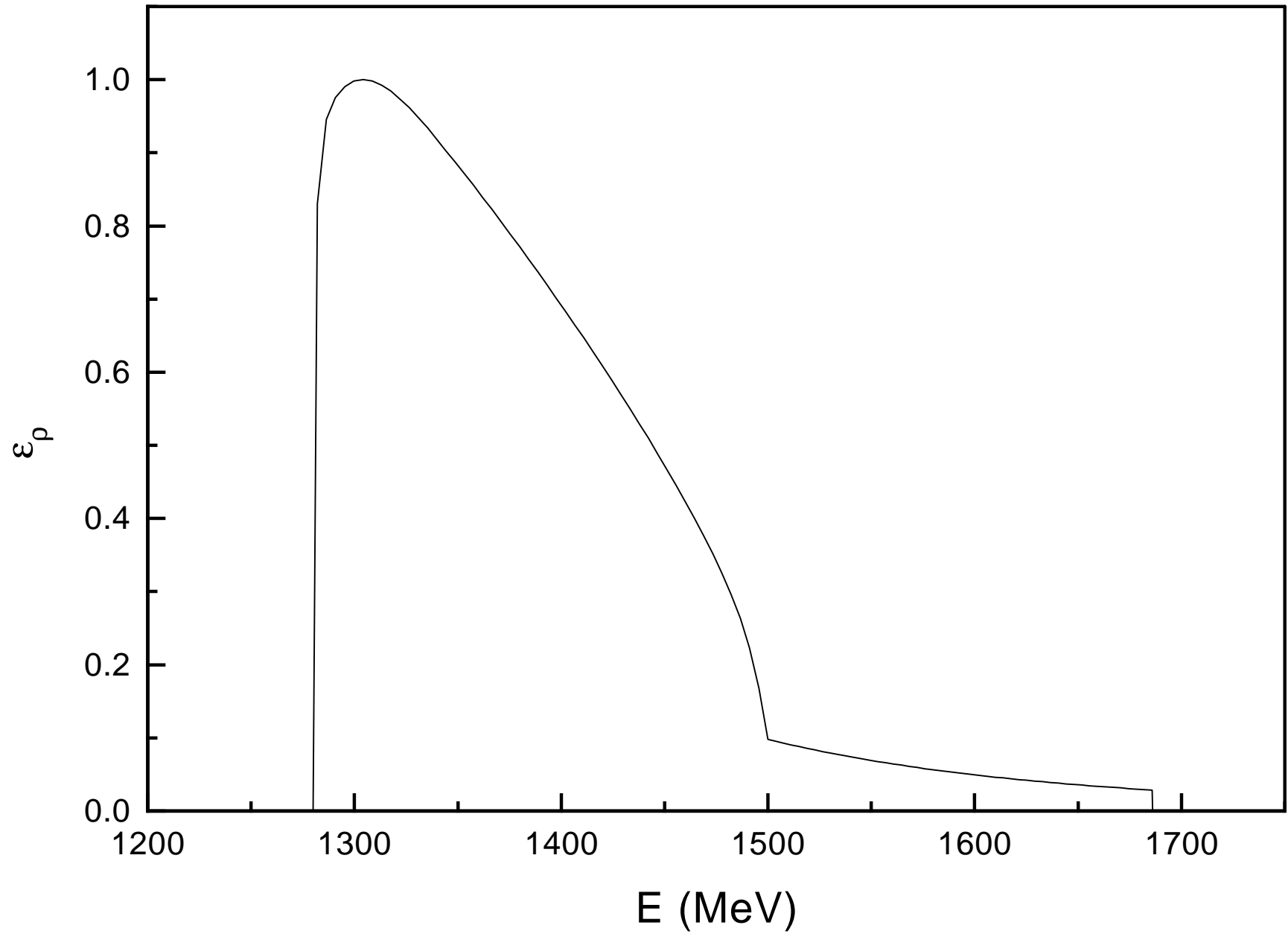
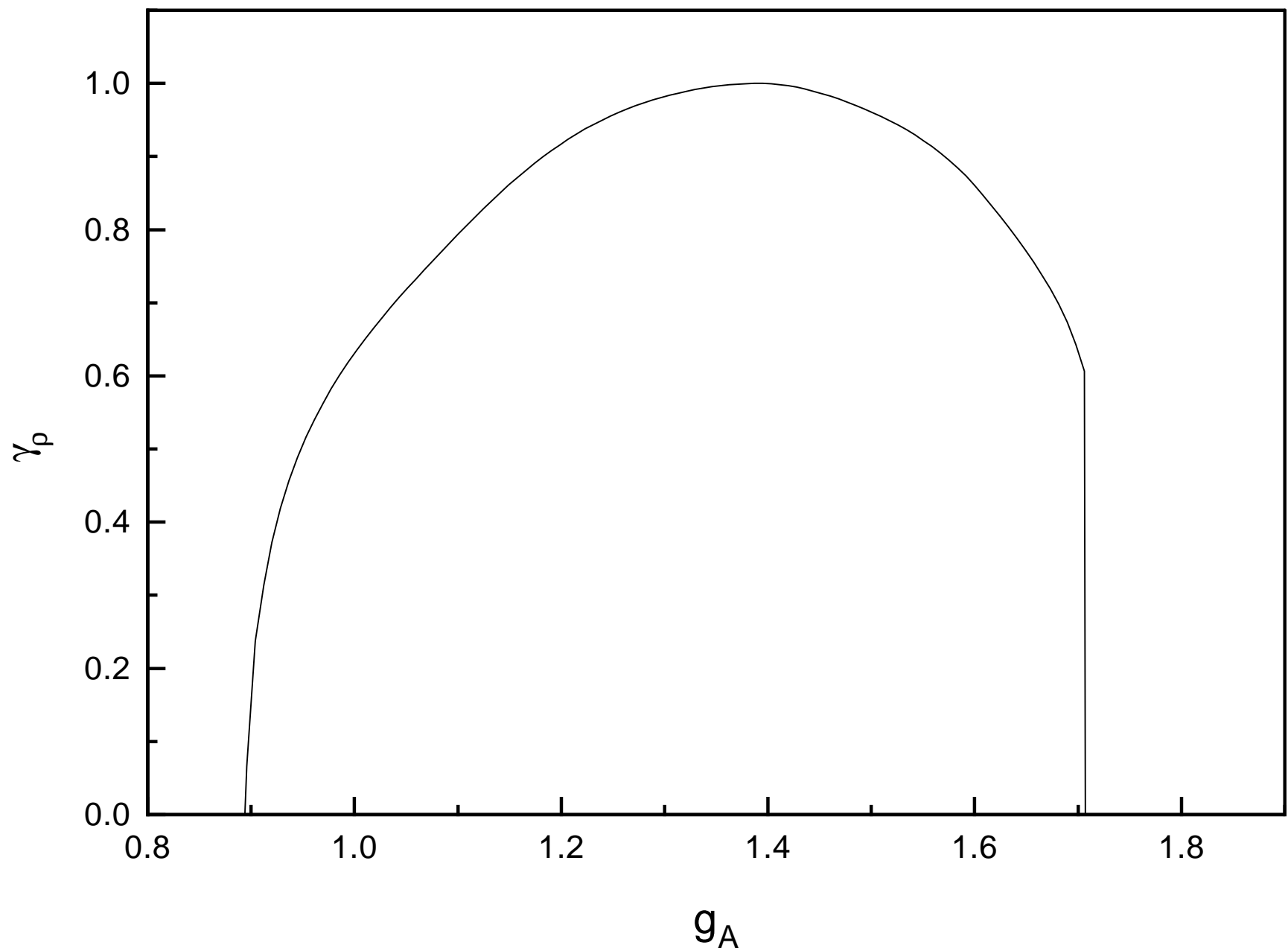
\end{document}